\documentclass[twocolumn]{aa} 

\usepackage{hyperref}
\usepackage{graphicx}
\usepackage{txfonts}
\usepackage{CJK}
\usepackage{booktabs}
\usepackage{graphbox}
\usepackage{amsmath}
\usepackage{soul}
\usepackage{scrextend}
\deffootnote[0.45em]{0.2em}{0.2em}{\textsuperscript{\thefootnotemark}\,}
\usepackage{amsmath,url}
\usepackage{algorithm}
\usepackage{amsfonts}
\usepackage{mathrsfs}
\usepackage{bm}
\usepackage{rotating}
\usepackage{color}
\usepackage{graphicx}
\usepackage{subfigure}
\usepackage[toc,page]{appendix}
\usepackage[english]{babel}
\usepackage[mathlines,switch]{lineno}
\newcommand\gaia{\textit{Gaia~}}
\newcommand\gdr[1]{\gaia~DR#1}

\newcommand\masyr{\ensuremath{\text{mas~yr}^{-1}}}

\newcommand\kms{\ensuremath{\text{km~s}^{-1}}}

\newcommand\gband{\ensuremath{G}}
\newcommand\gbp{\ensuremath{G_\mathrm{BP}}}
\newcommand\grp{\ensuremath{G_\mathrm{RP}}}
\newcommand\RAdeg{\ensuremath{\text{$\alpha$}}}
\newcommand\DEdeg{\ensuremath{\text{$\delta$}}}
\newcommand\plx{\ensuremath{\text{$\varpi$}}}
\newcommand\pmra{\ensuremath{\text{$\mu_{\alpha}^*$}}}
\newcommand\pmdec{\ensuremath{\text{$\mu_{\delta}$}}}

\newcommand\figref[1]{Fig.~\ref{#1}}
\newcommand\tabref[1]{Table~\ref{#1}}
\newcommand\secref[1]{Section~\ref{#1}}

\titlerunning{A new binary cluster catalog}
\authorrunning{Liu et al}
\begin{document} 
\begin{CJK*}{UTF8}{gbsn}
   \title{Binary clusters in the Galactic I: Systematic identification and classification using \gdr{3}}
    \author{Guimei Liu (刘桂梅)\inst{1,2}
          \and Yu Zhang (张余)\inst{1,2}
          \and Jing Zhong (钟靖)\inst{3}
          \and Songmei Qin (秦松梅)\inst{3,2,4}\\
          \and Yueyue Jiang (蒋悦悦)\inst{3,2,4}
          \and Li Chen (陈力)\inst{3,2}
          }
   \institute{XinJiang Astronomical Observatory, Chinese Academy of Sciences,150 Science 1-Street, Urumqi, Xinjiang 830011, China  \email{zhangyu@xao.ac.cn}
   \and School of Astronomy and Space Science, University of Chinese Academy of Sciences, No. 19A, Yuquan Road, Beijing 100049, China
    \and Astrophysics Division, Shanghai Astronomical Observatory, Chinese Academy of Sciences,80 Nandan Road, Shanghai 200030, China  \email{jzhong@shao.ac.cn}
    \and Institut de Ci\`encies del Cosmos, Universitat de Barcelona (ICCUB), Mart\'i i Franqu\`es 1, 08028 Barcelona, Spain}
   \date{Received ..., 2025; accepted ...}
\abstract
   {Binary clusters (BCs) provide valuable observational constraints on the formation, early evolution, and dynamical interactions of star clusters. Their spatial and kinematic associations offer unique insights into the hierarchical star formation process and tidal interactions within the Galactic disk.}
   {We aim to identify and classify BCs using high-precision astrometric and kinematic data, and to investigate their physical properties, mutual gravitational interactions, and formation rates.}
   {We used a comprehensive star cluster catalog that contains 4,084 high-quality clusters. Based on spatial and kinematic proximity, we identified 400 cluster pairs involving 686 unique clusters. These pairs were classified into three types: primordial BCs, systems formed through tidal capture or resonant trapping, and hyperbolic encounter pairs. For each system, we calculated the tidal factor to quantify the strength of mutual tidal interaction. 
   Additionally, we constructed multi-cluster systems by identifying transitive connections among cluster pairs.}
   {Among the 400 identified cluster pairs, nearly 60.8\% (243 pairs) are probably primordial BCs, exhibiting both similar ages and motions. This supports a scenario where they formed together in the same giant molecular cloud. We find that 82.5\% of the cluster pairs have strong mutual tidal forces.  In addition, 278 star clusters are identified as members of 82 multi-cluster systems, including 27 newly reported groups. Cross-matching with the literature confirms the recovery of previously reported systems and leads to the discovery of 268 new cluster pairs. In our sample, about 16.8\% of star clusters are involved in some type of interaction with another cluster, and 9.94\% of star clusters are likely born in primordial BCs.}
   {Our results provide a comprehensive, homogeneously identified sample of Galactic BCs. The high fraction of primordial BCs and their mutual tidal interaction suggest that cluster formation in pairs is a main outcome of star formation. This work offers new observational constraints on the formation and dynamical evolution of multiple star cluster systems.}
   \maketitle

\section{Introduction} \label{sec:intro}
Open clusters (OCs) are stellar systems consisting of dozens to thousands of stars that form from the same giant molecular cloud (GMC) and remain gravitationally bound \citep{Lada2003}. While most OCs exist as independent stellar systems, some form in groups, such as binary clusters (BCs) or more complex configurations involving multiple OCs \citep{Conrad2017, Bica2003A&A405991B, Camargo2016, Liu2025}. A well-known BC is the $h$ and $\chi$ Persei pair (NGC 869 and NGC 884; \citealt{Messow1913, Zhong2019A&A}), which illustrates the possibility of nearly coeval and physically associated clusters forming nearby.

BCs offer unique insights into both the star formation processes and subsequent dynamical evolution in different Galactic environments. Their existence challenges the classical theory that star clusters form and evolve in isolation. Instead, it supports the idea that star formation may occur in a clustered or hierarchical formation mode \citep{Bhatia1988, Dieball2002, Liu2025}, extending from small-scale associations to larger Galactic structures \citep{Elmegreen1996}.

The presence of BCs naturally raises questions about their origin. Several mechanisms have been proposed to explain the formation of interacting pairs \citep{delaFuenteMarcos2009a}, each leading to different observable properties. In the simultaneous formation scenario, both clusters form nearly simultaneously within the same GMC. As a result, they typically have similar ages, chemical compositions, and kinematics. These systems are considered primordial BCs. In contrast, sequential formation involves one cluster triggering the formation of another through stellar feedback \citep{Brown1995, Goodwin1997}, such as stellar winds or supernova explosions. Such pairs typically exhibit a small age difference and may or may not share similar kinematics, depending on whether they formed from the same GMC. Recent work by \citet{Liu2025} identified several primordial OC groups that likely formed through supernova-triggered sequential star formation, providing support for a hierarchical picture of clustered star formation regulated by stellar feedback within GMC environments.
Another possibility is tidal capture or resonant trapping, where initially unrelated clusters become dynamically associated through gravitational interactions or resonances in the Galactic potential. These systems exhibit similar motions but differ in age and composition \citep{van1996, Dehnen1998, Simone2004}. Therefore, they do not have a common origin. Lastly, some close cluster pairs may be optical doubles, which are only spatially proximate but kinematically unbound and unrelated; they may arise from projection effects or temporary flybys. 

While the formation of BCs can be attributed to several distinct mechanisms, discriminating among them observationally requires accurate identification of physically associated cluster pairs. To investigate BCs in greater depth, it is first necessary to observationally confirm cluster pairs or multiple systems that exhibit kinematic correlations. 
Detection of multiple systems has been an area of interest since the 1990s, with significant advancements in the past decade. This effort has led to the development of several catalogs, each covering specific subsets of clusters based on particular selection criteria \citep[e.g.,][]{delaFuenteMarcos2009a, Soubiran2019, Piecka21, Casado2021a, Song22, Qin2023ApJS, Palma2025}. In some instances, improved detection techniques and high-precision astrometric data have revealed that objects previously classified as single clusters are binary systems \citep[e.g.,][]{Camargo2021, Qin2025}, thereby underscoring the limitations of earlier surveys.
However, the variety in selection methods and datasets makes direct comparison between studies difficult.

Before the Gaia era, the limited precision of astrometric measurements made it a challenge to reliably identify member stars of OC, thereby rendering the detection and characterization of BCs extremely difficult. Early efforts, such as \citet{Subramaniam1995}, estimated that BCs accounted for approximately 8\% of OCs by applying a spatial separation threshold of 20 pc and identifying 18 candidate pairs. Later, \citet{delaFuenteMarcos2009a} adopted physical (rather than projected) separations, selecting pairs with distances below 30 pc as potential interacting systems. Their results suggested that at least 12\% of OCs in the solar neighborhood are gravitationally interacting with nearby companions. Moreover, they proposed that the fraction of BCs in the Milky Way is comparable to that observed in the Magellanic Clouds, and approximately 40\% of these binary systems may have a common origin. More recently, \citet{Conrad2017} were the first to utilize full 6D parameters (\RAdeg, \DEdeg,\pmra, \pmdec, \plx, RV) to systematically investigate the presence of BCs and OC groups. Using selection criteria of spatial separation threshold of 100 pc and velocity difference thresholds of 10 \kms or 20 \kms, they identified 14 pairs and 5 groups under the stricter threshold, and 31 pairs and 5 groups under the more relaxed condition.

The advent of \gaia data has dramatically transformed the study of OCs by providing high-precision astrometric data (\RAdeg, \DEdeg, \plx, \pmra, \pmdec) alongside multi-band photometric data like \gband, \gbp, and \grp\ \citep{gaiadr2,gaiaEdr3, GaiaDR3_2022}. These data have enabled the identification of BCs with unprecedented accuracy. In recent years, significant progress has been made through increasingly refined selection methods. 
These advancements include approaches based solely on spatial proximity \citep{Liu2019}, or spatial proximity combined with velocity consistency \citep{Soubiran2019, Casado2021a}, as well as the incorporation of age similarity \citep{Qin2023ApJS, Liu2025}. Some approaches rely on shared member stars \citep{Piecka21}, while others integrate spatial proximity, proper motion similarity, and age agreement \citep{Song22, Li2024}. In addition, tidal interaction indicators have recently been introduced as an auxiliary criterion for confirming BCs \citep{Palma2025}.
The application of these improved techniques has not only enhanced the reliability of BC identification but also provided valuable insight into their connection with larger cluster groupings.

Despite increasing attention in recent years, the identification of BCs remains challenging. A key limitation originates from inconsistencies in selection criteria and OC catalogs used across different studies, which complicates direct comparison and integration of results. To advance our understanding of their formation and evolution, a high-quality, large-sample, and methodologically uniform search for BCs and cluster groups is essential.
In this study, we independently identify new BCs and groups using a high-quality, large-sample catalog of OCs. Our method incorporates full 6D phase-space information, as well as the age difference between each cluster and its nearest neighbor. 

This paper is structured as follows:
We describe the data, including the sources of the OC samples and in \secref{sec: data}.
In \secref{sec: method}, we present the
procedures used to derive cluster parameters and the BC identification approach.
The results of the BC search are given in \secref{sec: results}.
In \secref{sec: dis}, we compare our findings with previous studies and discuss the identification of OC groups within our sample.
Finally, we summarize our conclusions in \secref{sec: summary}.

\begin{figure*}[!htbp]
    \centering
    \includegraphics[width=0.9\linewidth]{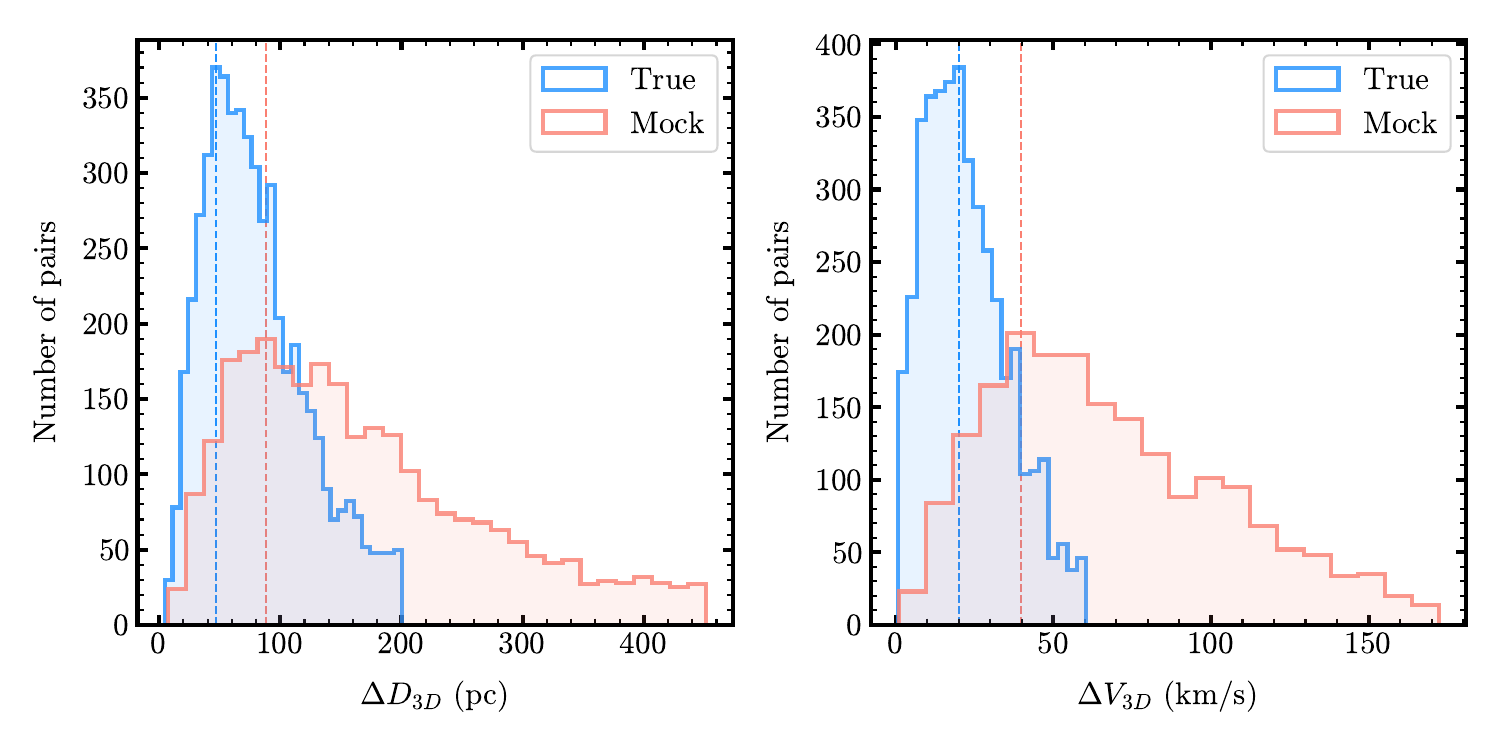}
    \caption{Comparison of the distributions for the spatial separation and velocity difference of the nearest neighbours. The salmon histogram is the result of one mock trial, while the blue histograms correspond to the results from our working sample.}
    \label{fig:typical_distribution}
\end{figure*}

\begin{figure*}[!htbp]
    \centering
    \includegraphics[width=0.9\linewidth]{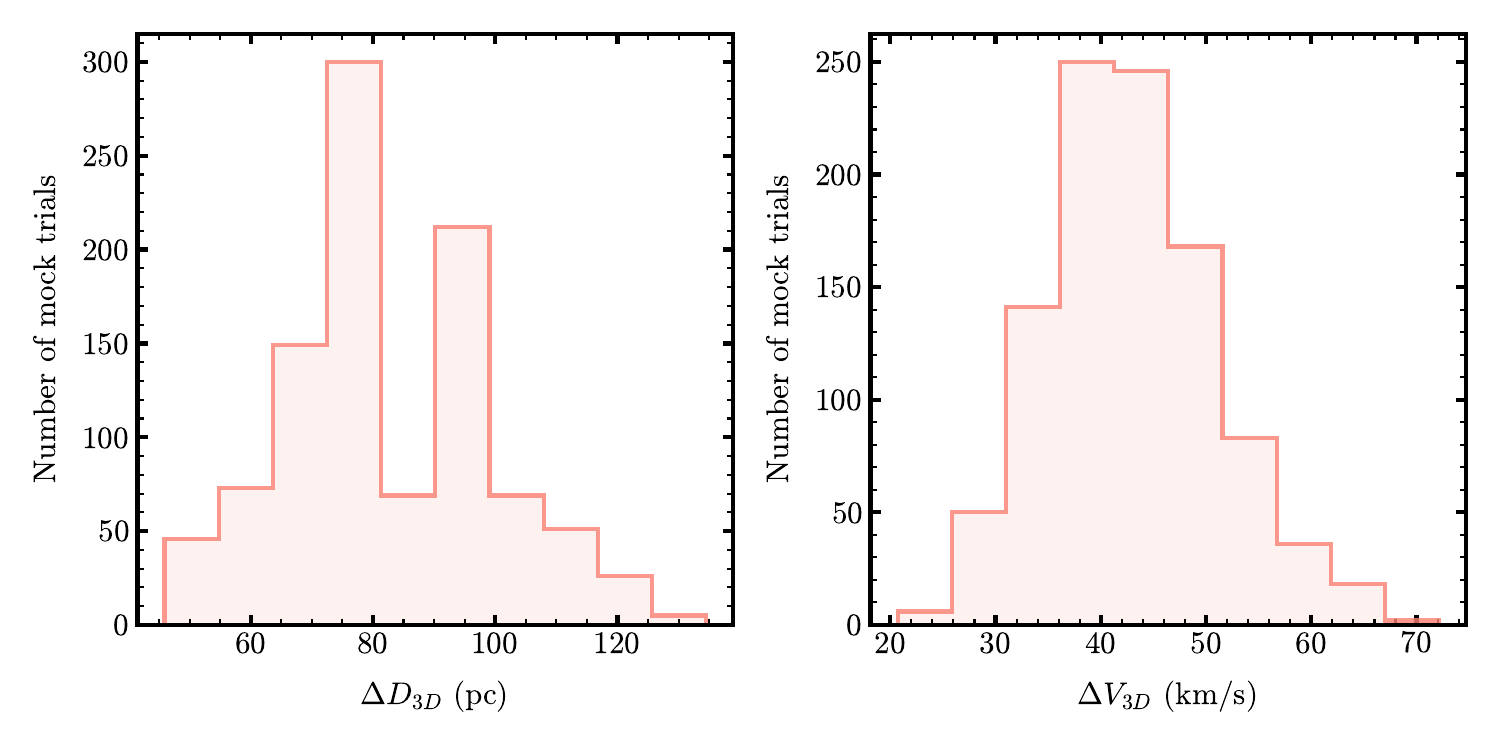}
    \caption{Distribution of peak values of spatial separation and velocity difference for nearest neighbours in 1000 mock trials.}
    \label{fig:mock_sep_Dv}
\end{figure*}

\begin{figure}[!htbp]
    \centering
    \includegraphics[width=0.98\linewidth]{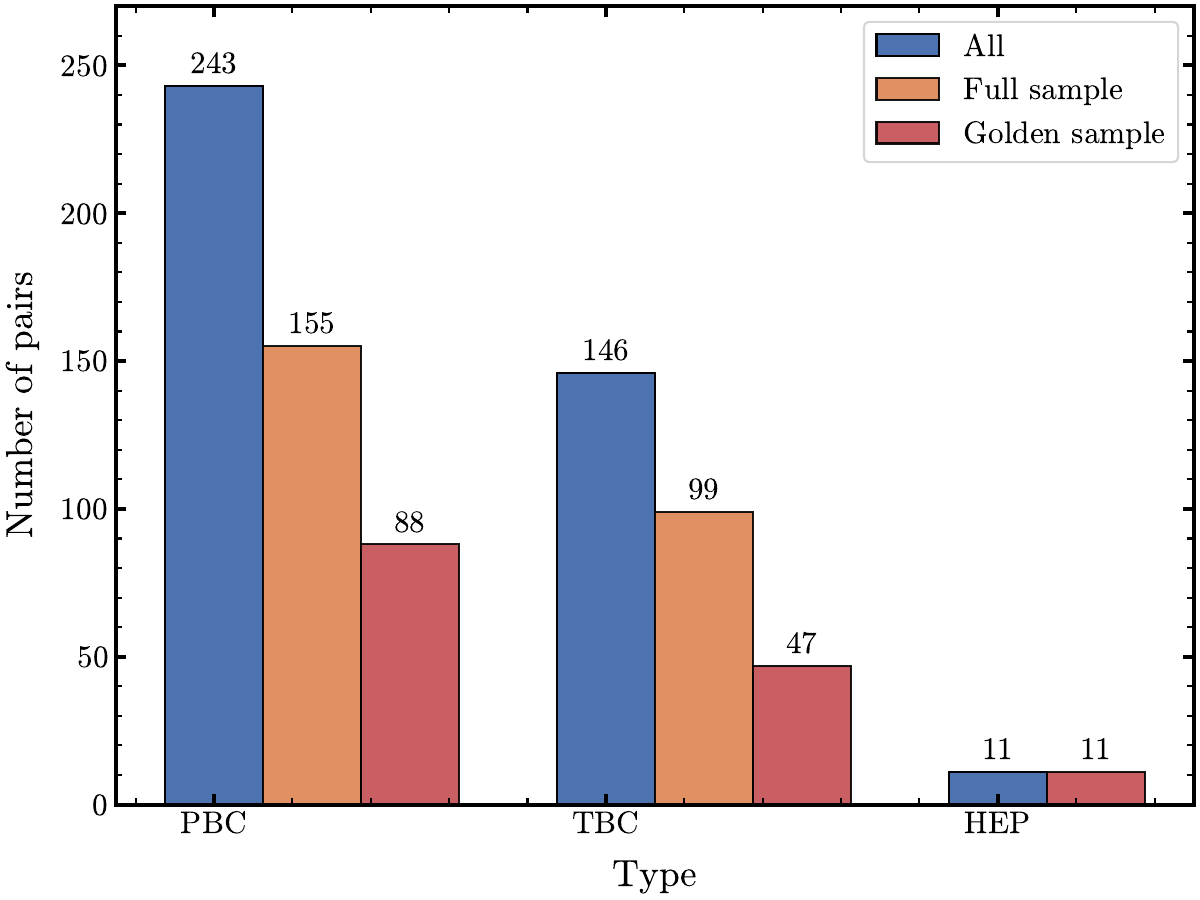}
    \caption{Distribution of different types of cluster pairs and sample category. The blue bars represent the total number of pairs in each type of cluster pairs, the orange bars indicate those in the full sample, and the red bars correspond to the golden sample. The numbers above each bar indicate the total number of pairs in that subcategory.Cluster pair types include primordial binary clusters (PBCs), tidal capture or resonant trapping binary clusters (TBCs), and Hyperbolic encounter pairs (HEPs), as defined in \secref{sec: classfy}}.
    \label{fig:classfy_pie}
\end{figure}

\begin{figure*}[!htbp]
    \centering
    \includegraphics[width=0.98\linewidth]{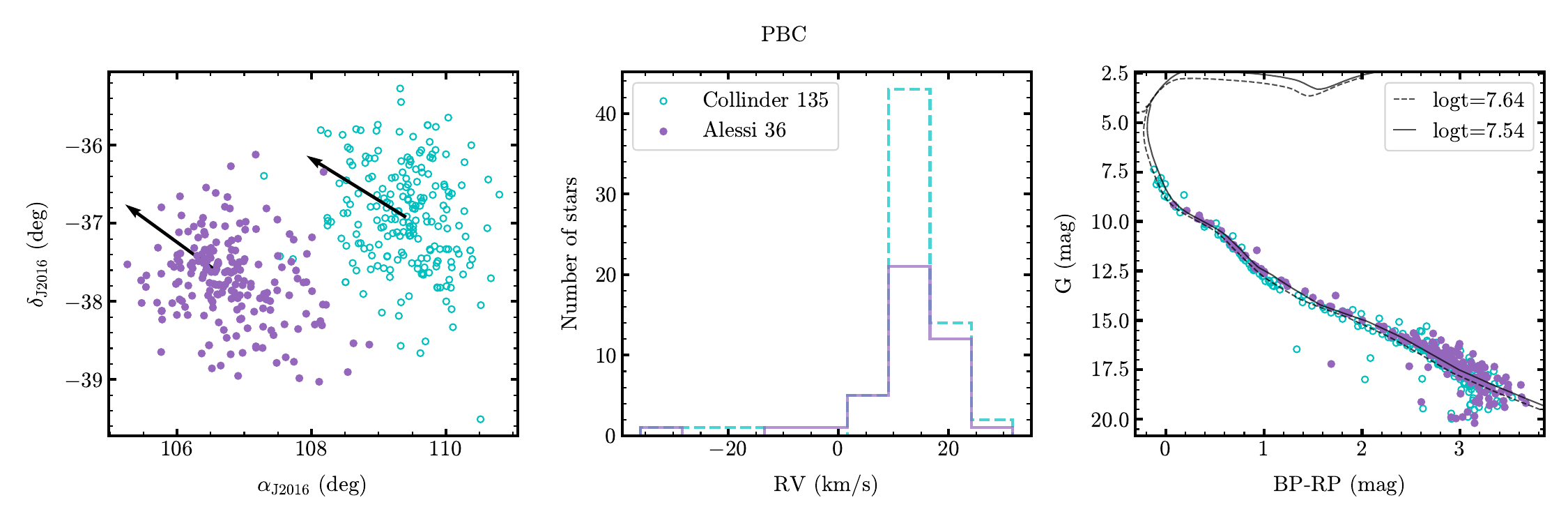}
    \includegraphics[width=0.98\linewidth]{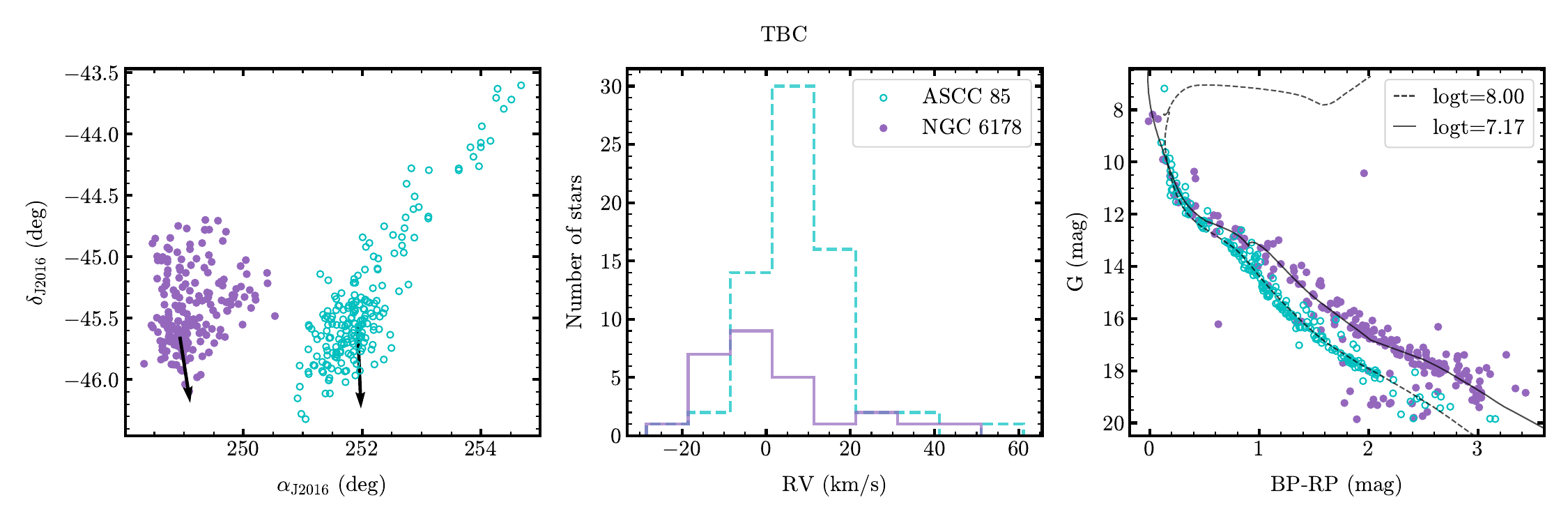}
    \includegraphics[width=0.98\linewidth]{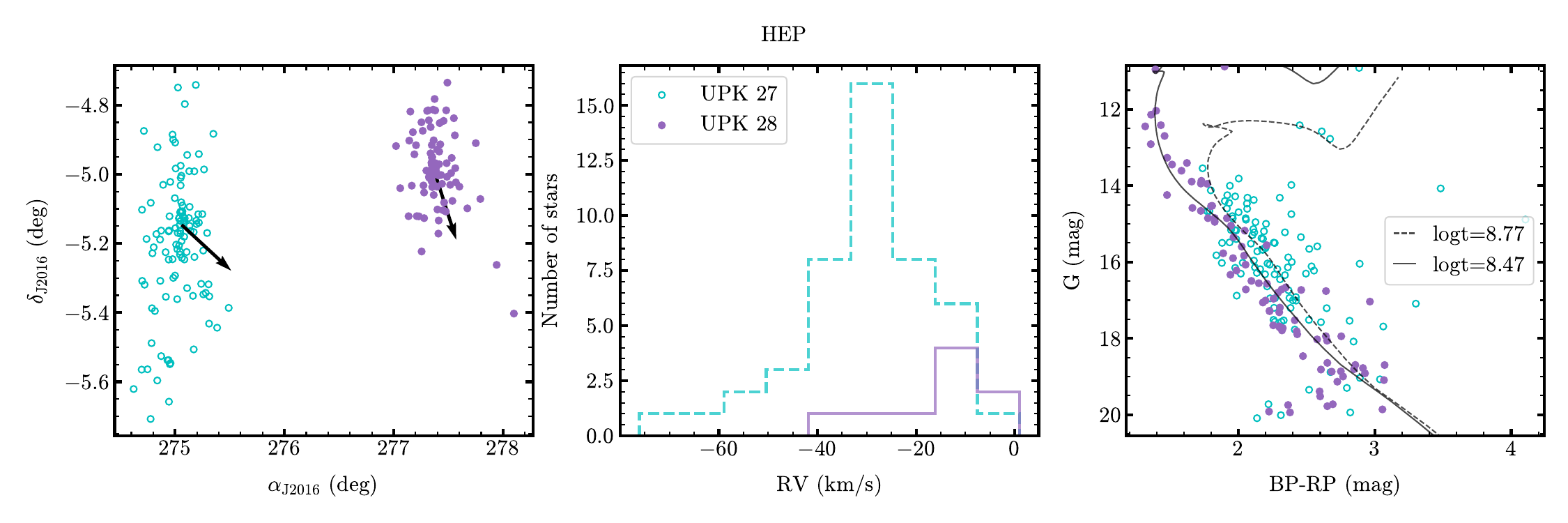}
    \caption{Multi-dimensional distributions of representative cluster pairs of different types. Shown are the distributions of member stars in spatial coordinates (left panel), radial velocity (middle panel), and color-magnitude diagram (CMD; right panel). Each point is colored according to the OC to which the star belongs. Arrows indicate the tangential velocities of the clusters, with each cluster pair plotted using a distinct velocity scale for clarity. }
    \label{fig: type3_bc}
\end{figure*}

\section{Data} \label{sec: data}
We search for BCs based on the homogeneous star cluster catalog published by \citealt{Hunt2024} (HR24 hereafter), which provides a comprehensive census of 7,167 star clusters. Among these, 4,782 are previously known star clusters compiled from literature, and 2,387 are newly identified. The cluster catalog provided by HR24 is an updated version of \citealt{Hunt2023} (HR23 hereafter), which was originally constructed from a systematic cluster search conducted in that study.
To detect and recover these star clusters, HR23 employed the HDBSCAN algorithm (Hierarchical Density-Based Spatial Clustering of Applications with Noise, \citealt{hdbscan_2013}). It is well-suited for identifying star clusters of varying densities and effectively separating member stars from field star contamination.

However, a major limitation of HDBSCAN is the high false-positive rate. This is due to the overconfidence of HDBSCAN, which sometimes misidentifies random statistical fluctuations in dense stellar fields as genuine star clusters. HR23 introduced the cluster significance test (CST) to mitigate this, which evaluates the statistical contrast between cluster members and the surrounding field stars \citep{Hunt2021}. Specifically, it compares the nearest-neighbor distance distribution of star cluster members with that of nearby field stars. Star clusters with CST values greater than 5$\sigma$ are considered to represent significant astrometric overdensities rather than random fluctuations, thus indicating genuine star clusters.

HR23 also assessed star cluster quality using the color-magnitude diagram (CMD), based on the assumption that star clusters typically consist of stars formed in the same GMC and thus share similar ages. This coeval population produces a well-defined sequence in the CMD. Thus, HR23 introduced a quality parameter called the CMD class, which is provided for each star cluster in the catalog. Star clusters with median CMD class values above 0.5 are considered to be consistent with a single stellar population and are thus more likely to be a real cluster. Combined with the CST, HR23 identified a reliable high-quality sub-sample of 4,105 clusters that satisfy CST greater than 5$\sigma$ and median CMD class greater than 0.5. 

Among these, HR23 further distinguished between gravitationally bound systems, such as OCs and globular clusters, and unbound systems that are more compatible with moving groups (MGs). In addition, HR23 derived a set of fundamental parameters for each star cluster, including key astrometric parameters such as position (\RAdeg, \DEdeg), proper motion (\pmra, \pmdec), parallax (\plx), and radial velocity (RV).

In our study, we selected a high-quality sample of 4,084 star clusters from a high-quality sub-sample provided by HR24, which includes both OCs and MGs. For each cluster, we use the full set of 6D phase-space parameters (\RAdeg,\DEdeg,\pmra,\pmdec,\plx, RV) provided by HR24, as well as the $R_{50}$, defined as the radius containing 50\% of the member stars.

\begin{table}
\centering
\caption{Compilation of multiple systems reported in the literature.}
\label{tab: literature}
\footnotesize 
\setlength{\tabcolsep}{2.5pt}
\renewcommand{\arraystretch}{1.5} 
\begin{tabular}{llc}
\hline
{Type}& {Authors}& {Number} \\
 \hline
 Binary & \citet{Subramaniam1995} & 18  \\
   { }& \citet{delaFuenteMarcos2009a} & 43   \\
     { }& \citet{Conrad2017} & 14   \\
     { } & \citet{Soubiran2019} & 8   \\
      { }& \citet{Liu2019} & 40   \\
      { }& \citet{Piecka21} & 50  \\
     { } & \citet{Casado2021a} & 11  \\
      { }& \citet{Casado2021b} & 1  \\
     { } & \citet{Angelo22} & 5 \\
    { }  & \citet{Song22} & 14  \\
    { }  & \citet{Qin2023ApJS} & 19  \\
    { }  & \citet{Li2024} & 13  \\
    { }  & \citet{Palma2025} &617 \\
\hline
Group & \citet{Pavlovskaya1989} & 5  \\
 { }&  \citet{Piskunov2006} &4\\
 { }&  \citet{Conrad2017}  & 5  \\
 { }&  \citet{Soubiran2019}  & 4  \\
 { }&  \citet{Liu2019}  & 15  \\
  { }&  \citet{Cantat2019ring}  & 1  \\
  { }&  \citet{Beccari2020}  & 1  \\
  { }&  \citet{Tian2020}  & 1  \\
  { }&  \citet{Pang2021}  & 1  \\
 { }&  \citet{Piecka21}  & 10 \\
   { }& \citet{Casado2021a}  & 11 \\
     { }&  \citet{Pang2022}  & 4  \\
  { }&  \citet{Angelo22}  & 2 \\
{ }&  \citet{Kounkel2022} &1\\
 { }&\citet{Qin2023ApJS} & 3  \\
{ }&  \citet{Palma2025} &261 \\
 { }&  \citet{Liu2025} &4\\
 \hline
\end{tabular}
\end{table}

\section{Identification criteria} \label{sec: method}
We derived the parameters used in the identification process by computing the 3D positions, 3D velocities, and tangential velocities ($V_t$) for all star clusters. $V_t$ are obtained by converting proper motion measurements using the following equation:
\begin{equation}
    \centering
    V_t = k \times d \times PM,
    \label{equ:vt}
\end{equation}
where $V_t$ is the tangential velocity in \kms, $k=4.74$ is the conversion factor to ensure unit consistency, PM is the proper motion component in \masyr\ and $d$ is the heliocentric distance of the clusters in kpc.
To determine the closest neighbor of each OC in 3D space,  we converted the equatorial coordinates to Galactocentric Cartesian coordinates using the \texttt{Astropy} package \citep{Astropy2013A&A, Astropy2018AJ}. For the transformation, we adopted the position of Sun as ($X, Y, Z$) = [–8, 0, 0.015] kpc and velocity as ($U, V, W$) = [10, 235, 7] \kms\ in the Galactocentric reference frame \citep{Bovy2015}.

BCs are generally expected to show close spatial proximity and similar kinematics. Consequently, compared to random cluster pairs, their spatial separations and velocity differences should be significantly smaller than those of randomly associated cluster pairs \citep{Conrad2017}. We first investigate the statistical properties of spatial and kinematic proximity among OCs to establish reliable thresholds for identifying BCs.

To quantify this, we began with a sample of 4,084 high-quality star clusters and performed a nearest-neighbor analysis in 3D space ($X, Y, Z$). For each star cluster, we identified its closest neighboring cluster and calculated the 3D spatial separation ($\Delta D_{3D}$) between them. We also computed their relative velocity difference, including the tangential velocity difference ($\Delta V_t$) and also full 3D velocity difference ($\Delta V_{3D}$). After removing duplicate pairs from the initial set of 4,084 pairs, we obtain a final sample of 2,974 unique cluster pairs. The distributions of $\Delta D_{3D}$ and $\Delta V_{3D}$ of 2,974 pairs are shown in \figref{fig:typical_distribution}. The $\Delta D_{3D}$ distribution exhibits a prominent peak at 50 pc, while the $\Delta V_{3D}$ distribution peaks at approximately 20 \kms. These peak values represent the typical characteristics of physically associated cluster pairs. 

To test the robustness of $\Delta D_{3D}$ and $\Delta V_{3D}$ statistical distribution, we constructed a mock dataset by randomly assigning the 6D parameters (\RAdeg,\DEdeg,\pmra,\pmdec,\plx, RV) from real star clusters to generate artificial mock clusters. Specifically, for each mock star cluster, we independently sampled each parameter from the distributions of the real 4,084 star clusters, thereby preserving the marginal distributions but destroying any intrinsic correlations. This process yielded a mock catalog of 4,084 star clusters.

We then performed a nearest-neighbor search for each mock OCs and calculated the $\Delta D_{3D}$ and $\Delta V_{3D}$ between each mock pair. This procedure was repeated over 1,000 independent trials, each using a newly generated mock catalog with randomized parameters. \figref{fig:typical_distribution} presents the results of one of these trials, where the distribution of $\Delta D_{3D}$ and $\Delta V_{3D}$ of the mock dataset is significantly broader than that of the real data. While \figref{fig:mock_sep_Dv} shows the distribution of peak values from the entire set of 1,000 trials, the mock dataset yields a typical $\Delta D_{3D}$ of approximately 80 pc between nearest-neighbor pairs, and a typical $\Delta V_{3D}$ of about 40 \kms.
Both values are significantly higher than the peaks observed in the real data. The larger values obtained from the simulations are likely a result of the random assignment disrupting the original kinematic coherence present in the real data, which is attributed to gravitational interaction. This leads to a more uniform distribution of positions and velocities, producing greater separations and velocity differences in the absence of physical associations.
In contrast, the tighter distribution observed in the real sample indicates an intrinsic connection between OC pairs. This result demonstrates high statistical significance and confirms that $\Delta D_{3D}$ below 50 pc and a $\Delta V_{3D}$ below 20 \kms\ are meaningful thresholds likely corresponding to genuine BC systems rather than chance alignments. This result supports the use of $\Delta D_{3D}$ threshold of 50 pc and a $\Delta V_{3D}$ threshold of 20~\kms\ as effective criteria for identifying physically bound or primordial BC systems.

The identification process benefits from the inclusion of the $\Delta V_t$, which constrains the alignment of velocity vectors between OCs and strengthens the detection of systems with coherent motion. Since nearly 27\% of OCs in the high-quality sample lack sufficient RV data (i.e., $N_{\rm RV} \leq 1$), which makes it difficult to compute accurate $\Delta V_{3D}$ for all candidate pairs. To address this limitation, the $V_t$ component serves as an alternative kinematic indicator.Following previous studies \citep[e.g.,][]{Conrad2017,Qin2023ApJS,Liu2025}, we applied an empirical threshold of $\Delta V_t < 10$ \kms. 

\begin{table*}[htbp]
\centering
\caption{Binary cluster candidates.}
\setlength{\tabcolsep}{8.0pt}
\renewcommand{\arraystretch}{1.5}
\begin{tabular}{rrrrrrrrrrrr}
\hline 
Pair & $\text{ID}_\text{HR24}$ & Cluster &$\Delta D_{3D}$ & $\Delta V_{t}$  & $\Delta V_{3D}$ & $\Delta$Age & logt & DM & $E(B-V)$ &Type& overlap  \\ 
{} & {} & {} &pc & \kms & \kms & Myr & {} & mag &mag &{}&{}\\
(1) & (2) & (3) & (4) & (5) & (6) & (7) & (8) & (9) & (10) & (11) & (12)  \\
\hline 
1 & 5514 & Theia\_117 & 39.25 & 6.53 & 9.63 & 17.80 & 7.74 & 7.19 & 0.12 & PBC & -- \\
1 & 1 & ADS\_16795 & 39.25 & 6.53 & 9.63 & 17.80 & 7.57 & 7.06 & 0.13 & PBC & -- \\
2 & 7 & ASCC\_12 & 43.21 & 0.91 & 16.15 & 69.23 & 8.02 & 11.09 & 0.33 & TBC & --  \\
2 & 4203 & HXHWL\_18 & 43.21 & 0.91 & 16.15 & 69.23 & 7.55 & 10.68 & 0.31 & TBC &--   \\
3 & 481 & CWNU\_350 & 41.45 & 0.49 & 5.20 & 58.50 & 8.05 & 10.20 & 0.19 & TBC & --  \\
3 & 8 & ASCC\_13 & 41.45 & 0.49 & 5.20 & 58.50 & 7.73 & 10.80 & 0.25 & TBC & --  \\
4 & 5481 & Theia\_13 & 20.07 & 3.11 & 4.14 & 6.93 & 7.13 & 8.16 & 0.09 & PBC &--   \\
4 & 10 & ASCC\_18 & 20.07 & 3.11 & 4.14 & 6.93 & 7.31 & 8.09 & 0.13 & PBC & --  \\
5 & 6925 & UBC\_17a & 17.92 & 1.12 & 1.68 & 9.72 & 6.50 & 8.58 & 0.13 & PBC & --  \\
5 & 11 & ASCC\_19 & 17.92 & 1.12 & 1.68 & 9.72 & 7.11 & 7.91 & 0.11 & PBC &--  \\
... &... & ... & ... &... &... & ... & ...  & ... &... & ... & ...  \\ 
\hline
\end{tabular}
\tablefoot{Column (1): Pair IDs. Column (2): Cluster IDs from \citealt{Hunt2024}. Column (3): Star cluster names. Column (4): The 3D separation between clusters, derived from \RAdeg, \DEdeg, and \plx\ at the epoch J2016.0. Column (5): The tangential velocity differences between clusters. Column (6):  The 3D velocity differences between clusters.  Column (7):  Age differences of the pair clusters. Column (8): Cluster age determined by the isochrone fit. Column (9): Distance modulus determined by the isochrone fit. Column (10): Reddening determined by the isochrone fit. Column (11): Pair classification. Column (12): Previous references.
The full Table is available at the CDS. Here we only show the first five BCs.}
\label{tab: bc_candidate}
\end{table*}

\section{Results} \label{sec: results}
\subsection{Identification of BC candidates}
We first selected cluster pairs with $\Delta D_{3D} < 50$ pc and $\Delta V_t < 10$ \kms. This initial selection yielded a sample of 400 candidate BCs with close proximity and coherent tangential velocity.
To further assess the reliability of these candidate BCs, we evaluated the availability of RV data for both member clusters. Based on this, we divided the 400 candidates into two categories:
\\
$-$ Golden Sample: Pairs in which both member star clusters have a sufficient number of stars with measured radial velocities. Specifically, we require that each cluster in the pair has at least ten member stars with available RV measurements. Pairs failing to meet this criterion are excluded from the golden sample.
\\
$-$ Full Sample: Pairs that satisfy the $\Delta D_{3D}$ and $\Delta V_t$ thresholds, but lack sufficient RV data in at least one member OC (i.e., fewer than ten stars with RVs). Although $\Delta V_{3D}$ cannot be reliably computed for these systems, their spatial proximity and aligned tangential velocity suggest a likely physical association. 

Based on this criterion, we identify 146 pairs as the high-quality golden sample and 254 pairs as the full sample, which is shown in \figref{fig:classfy_pie}. This division enables a distinction between robust, fully characterized BCs and those that remain plausible but less certain due to incomplete velocity information.

\subsection{Classification of BCs}\label{sec: classfy}
To further classify the 400 candidate BCs, we considered the age difference ($\Delta$Age) between the member OCs. The member OCs typically have similar ages for BCs with a common origin, while sequentially formed pairs may exhibit a small but non-negligible age difference. If the secondary cluster forms within the same molecular cloud as the primary one, the two OCs are expected to share similar kinematic properties. In such scenarios, the pair can still be considered to have a common origin and classified as a primordial BC system.

Previous studies have proposed different thresholds for the $\Delta$Age between member OCs of BC. For example, \citet{delaFuenteMarcos2009a} adopted a limit of 50 Myr, while \citet{Qin2023ApJS,Liu2025} used a stricter criterion of 30 Myr. Following this precedent, we adopt 30 Myr as the criterion for identifying cluster pairs with a common origin.
Based on these analyses, we classified the BC candidates into the following categories:
\begin{enumerate}
    \item Primordial binary cluster (PBC): The member OCs have similar ages and share common kinematics, consistent with formation from the same GMC. In this study, PBC is defined using the following criteria:
    \begin{itemize}
        \item $\Delta V_{3D} \leq 20$ \kms
        \item $\Delta \mathrm{Age} \leq 30$ Myr
    \end{itemize}
    These criteria ensure that the member OCs are not only kinematically associated but also likely coeval, supporting a common origin scenario.

    \item Tidal capture or resonant trapping binary cluster (TBC): These cluster pairs occupy a limited volume of space and exhibit similar kinematics, but their ages differ significantly. The dynamical association is likely due to gravitational interactions such as tidal capture or resonant trapping rather than a common formation history. The classification is based on:
    \begin{itemize}
        \item $\Delta V_{3D} \leq 20$ \kms
        \item $\Delta \mathrm{Age} > 30$ Myr
    \end{itemize}

    \item Hyperbolic encounter pair (HEP): These are cluster pairs that are spatially close but have significantly different velocities, indicating they are not born gravitationally bound. Other physical properties such as age or metallicity may also differ. These are characterized by:
    \begin{itemize}
        \item $\Delta V_{3D} > 20$ \kms
    \end{itemize}
\end{enumerate}

To apply this classification scheme, we performed isochrone fitting for 400 pairs of BC candidates to obtain accurate age estimates.
We performed isochrone fitting on the CMD of each target OC, using theoretical isochrones from the \gaia photometric system provided by CMD 3.7 \footnote{\url{http://stev.oapd.inaf.it/cgi-bin/cmd}} \citep{Bressan2012, Chen2015, Marigo2017ApJ83577M}. The best-fit isochrone was determined through visual inspection of a series of isochrone fits spanning a logarithmic age range of log(Age[yr]) = 6 to 10, with intervals of 0.01, and a metallicity value of solar metallicity ($Z_\odot$=0.0152, \citealt{Caffau2011SoPh268255C}). 

Based on the classification approach described above, we classify the 400 candidate BCs into three types. \figref{fig: type3_bc} presents representative examples of the three types.
In both the PBC and TBC examples, the member OCs are spatially close and kinematically similar. The key distinguishing factor is their age. The member OCs of PBC have nearly identical ages, while members of TBCs show clear age differences. By contrast, HEPs exhibit large RV differences, indicating they are not coevally associated despite their spatial proximity.

The classification results are presented in \figref{fig:classfy_pie}. A total of 243 pairs are identified as PBCs, accounting for approximately 60.8\% of all BC candidates. This high proportion suggests that the majority of BC systems are likely to have formed together. Among them, 88 pairs belong to the golden sample, while 155 pairs come from the full sample.
In addition, 146 pairs are classified as TBCs, including 47 golden sample pairs and 99 full sample pairs. 
The properties of all BC systems are summarized in \tabref{tab: bc_candidate}.  
\begin{figure}[htbp]
    \centering
    \includegraphics[width=1\linewidth]{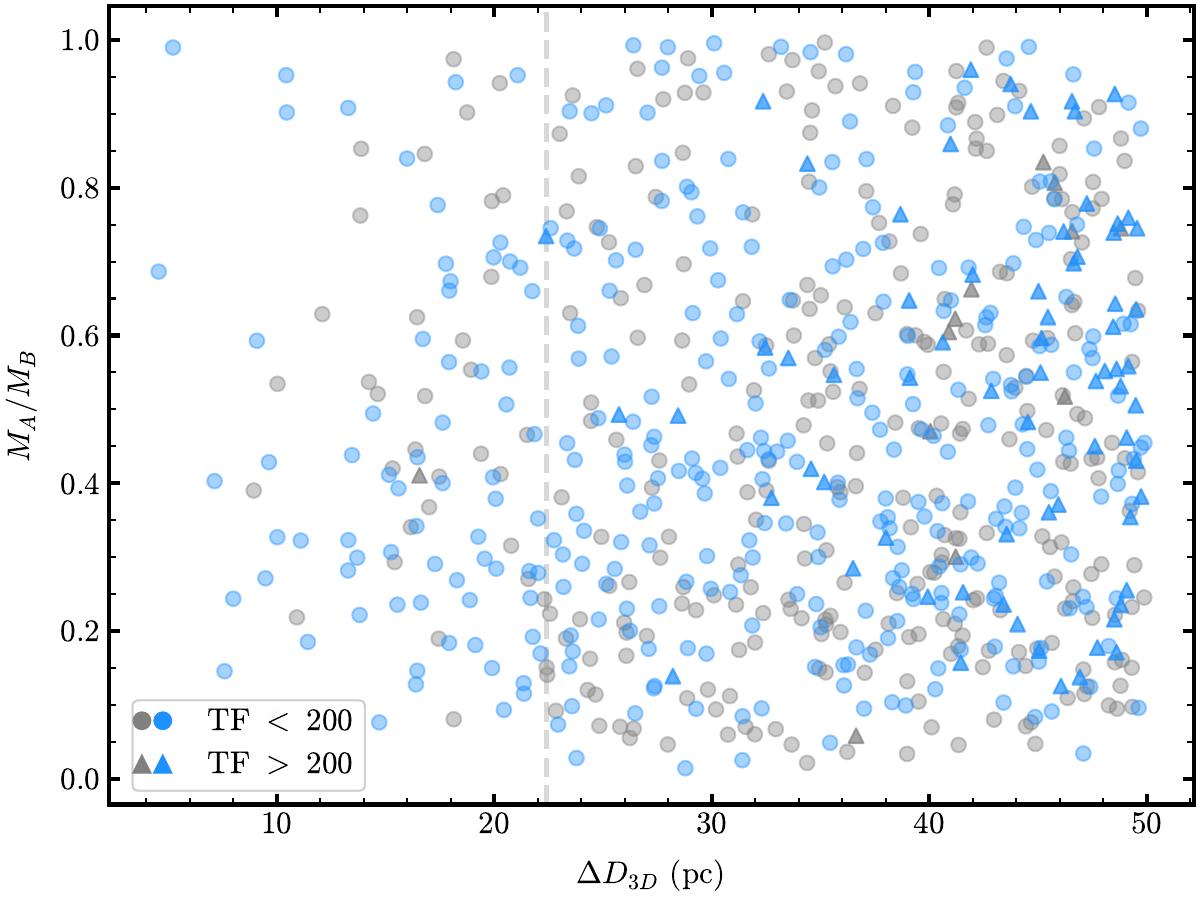}
    \caption{Mass ratio versus spatial separation for pairs with $\Delta D_{3D} \leq$ 50. Gray symbols represent pairs from the initial sample, while blue symbols represent cluster pairs identified in this work. Circles indicate pairs with TF values less than 200, and triangles indicate those with TF values greater than 200.}
    \label{fig: TF_sep}
\end{figure}

\section{Disscussion} \label{sec: dis}
\subsection{Tidal force}
We estimate the mutual tidal force between members in each pair using the tidal factor (TF), following the method of \citealt{Palma2025} (hereafter P25), where TF serves as a reciprocal proxy for the strength of the tidal force. Since the TF serves as the reciprocal proxy of the tidal force, each pair is assigned two TF values, and a lower TF value indicates a stronger tidal force exerted by the companion cluster.
The TF is calculated as follows:
\begin{equation}
\centering
    TF = \frac{d^3}{M_{cl}R_{50}},
\end{equation}
where $d$ is the 3D separation between the members in units of pc, $R_{50}$ is the half-member radius of the target OC in pc, and $M_{cl}$ is the mass of the companion OC in solar masses. To obtain the mass of each OC, we determined the nearest point on the best-fit isochrone (\secref{sec: classfy}) and assigned the corresponding mass as the star mass. The total OC mass was calculated by summing the masses of all member stars. 

By computing the TF distribution for previously reported BC samples, P25 found that most cluster pairs classified as binary or group systems have spatial separations concentrated around 50-100 pc, and TF values typically below $\sim$250. 
Based on this, P25 proposed TF$< 200$ as a threshold for identifying BCs.
To quantify the tidal interaction between BCs, we also used the threshold TF $<$ 200 to signify a significant tidal force between clusters.
Specifically, a pair is considered to exhibit tidal interaction when at least one cluster in a pair has a TF below this threshold. 

Using the TF metric, we found that the vast majority of identified BCs exhibit pronounced tidal interactions. Out of 400 candidate BCs in our sample, 330 pairs (82.5\%) have TF $<$ 200, indicating a significantly high tidal force between the components. 
Among them, 207 pairs are classified as PBCs, 114 as TBCs, and 9 as HEPs. 
The prevalence of low TF values among PBCs and TBCs strongly suggests that these systems are physically interacting: their mutual tidal forces are high, consistent with the expectation that they are either born together or bound by gravity. In such pairs, the gravity from the companion OC is actively influencing the other OC, for example, by tidally distorting the outer regions or facilitating mass exchange between the clusters. The independently identified BCs in this study are largely validated as physically interacting pairs under the tidal factor criterion proposed by P25, either through a common origin or dynamical capture.
Even in the few HEP systems that show TF $<$ 200, the low TF implies a transient but significant tidal effect.  In these cases, even without strong kinematic alignment, spatial proximity may still lead to a transient tidal force. This is an important distinction: HEPs are not long-term bound binaries, yet a sufficiently close approach can temporarily mimic the tidal influence seen in true bound pairs. Thus, a low TF can occur not only in enduring BC systems but also during short-lived encounters, highlighting that strong tidal forces can arise whenever two clusters come within a critical range, even if they will eventually separate. 

To gain deeper insight into the physical conditions governing tidal interactions, the relationship between the tidal factor, spatial separation, and mass ratio of cluster pairs was investigated, as these parameters jointly affect the strength of gravitational interaction. \figref{fig: TF_sep} presents a comparison of the BCs identified in this work with and without significant tidal interaction (defined as TF $<$ 200), illustrating how tidal effects vary with spatial separation and relative OC mass. Notably, no clear correlation is found between the tidal factor and the mass ratio, suggesting that the presence of tidal interaction is primarily governed by spatial proximity rather than the relative mass of the OC. 
A distinct boundary is observed in which nearly all BCs with $\Delta D_{3D} < 22$ pc exhibit TF values below 200, indicating uniformly strong mutual gravitational influence. This suggests that such close pairs are almost inevitably subject to significant tidal interactions, with forces strong enough to bind or substantially perturb one another. 
For cluster pairs with $\Delta D_{3D} \gtrsim 22$ pc, a mixture of tidal environments is observed. Some pairs still exhibit TF $<$ 200, whereas an increasing fraction display higher TF values indicative of weaker mutual interactions. This suggests that beyond this separation, spatial proximity alone is insufficient to guarantee strong tidal coupling between clusters.

To test the generality of this distance-dependent trend, we extended the analysis to all 716 cluster pairs with $\Delta D_{3D} < 50$ pc, including both the pairs identified as BCs in this work and those not selected. For the 316 non-selected pairs, we calculated the TF using cluster masses from the and \citealt{Hunt2024} (HR24 hereafter). This extended sample allowed us to construct a complete TF distribution for all close pairs. 
As shown in \figref{fig: TF_sep}, nearly all cluster pairs with $\Delta D_{3D} <22$ pc yielded TF $<$ 200, reinforcing the idea that at sufficiently small separations, significant tidal forces are almost ubiquitous.
This clear dependence on separation indicates that tidal forces become stronger as clusters move closer together, such that nearly all pairs with $\Delta D_{3D} < 22$ pc experience significant tidal interaction. These findings further justify the adoption of TF $<$ 200 as a physically motivated threshold, effectively identifying pairs that lie within the critical regime of mutual gravitational influence.

Potential uncertainties in the TF calculation should be acknowledged in this work. The OC masses used in our analysis are derived from robust isochrone fits. However, unresolved binary stars and the missing faint member stars are not fully accounted for when summing member masses. As a result, TF values are likely conservative, which means the true tidal factors could be slightly overestimated (underestimated tidal force). Consequently, some cluster pairs marginally above the threshold could satisfy TF $<$ 200 if better mass estimates were available. Overall, these uncertainties do not affect the general conclusion that the majority of close cluster pairs exhibit significant tidal interaction.
\begin{figure}[htbp]
    \centering
    \includegraphics[width=0.98\linewidth]{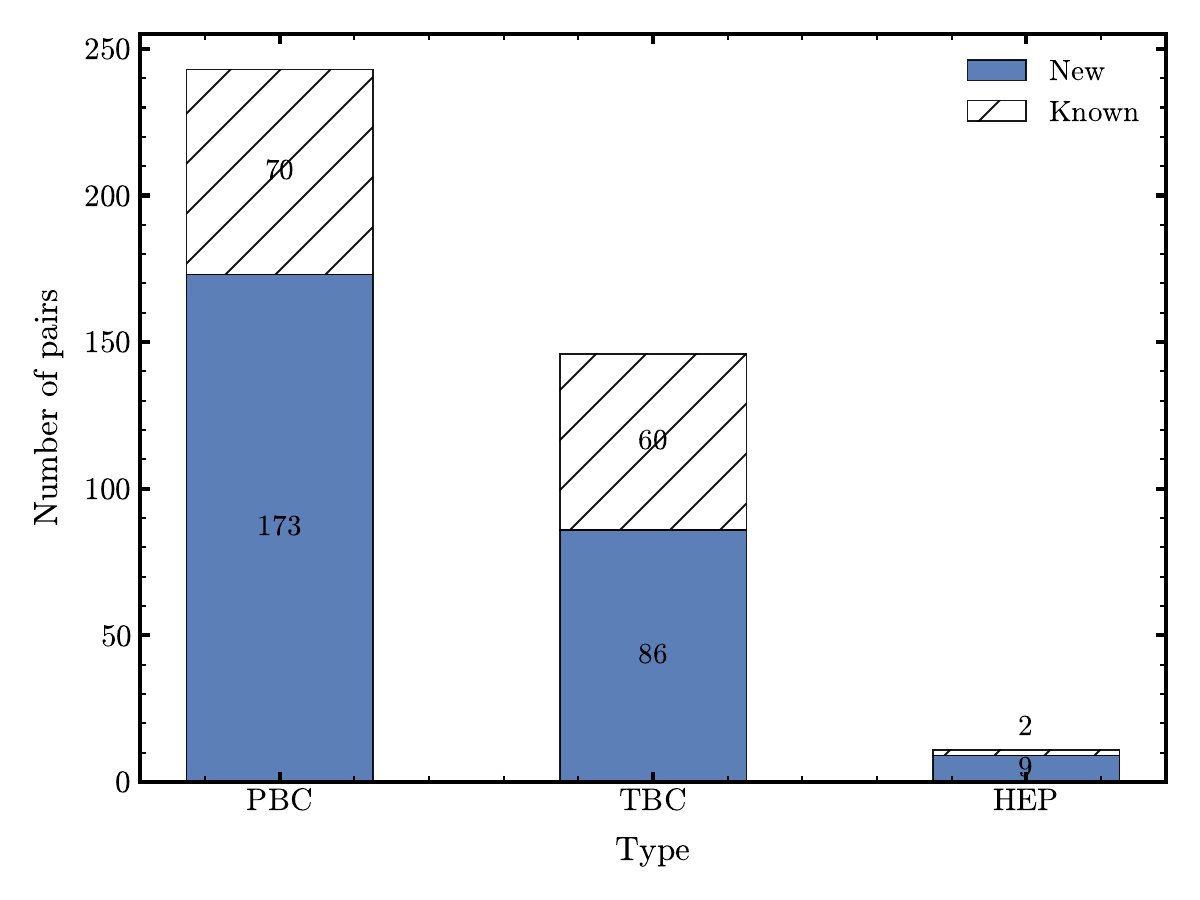}
    \caption{Cross-matching results with previous studies. Distribution of BC candidates by classification type (PBC, TBC, HEP), showing the number of systems that are new (blue) versus those previously reported in the literature (filled).}
    \label{fig: lit_repeat}
\end{figure}

\begin{figure}[!bp]
    \centering
    \includegraphics[width=0.98\linewidth]{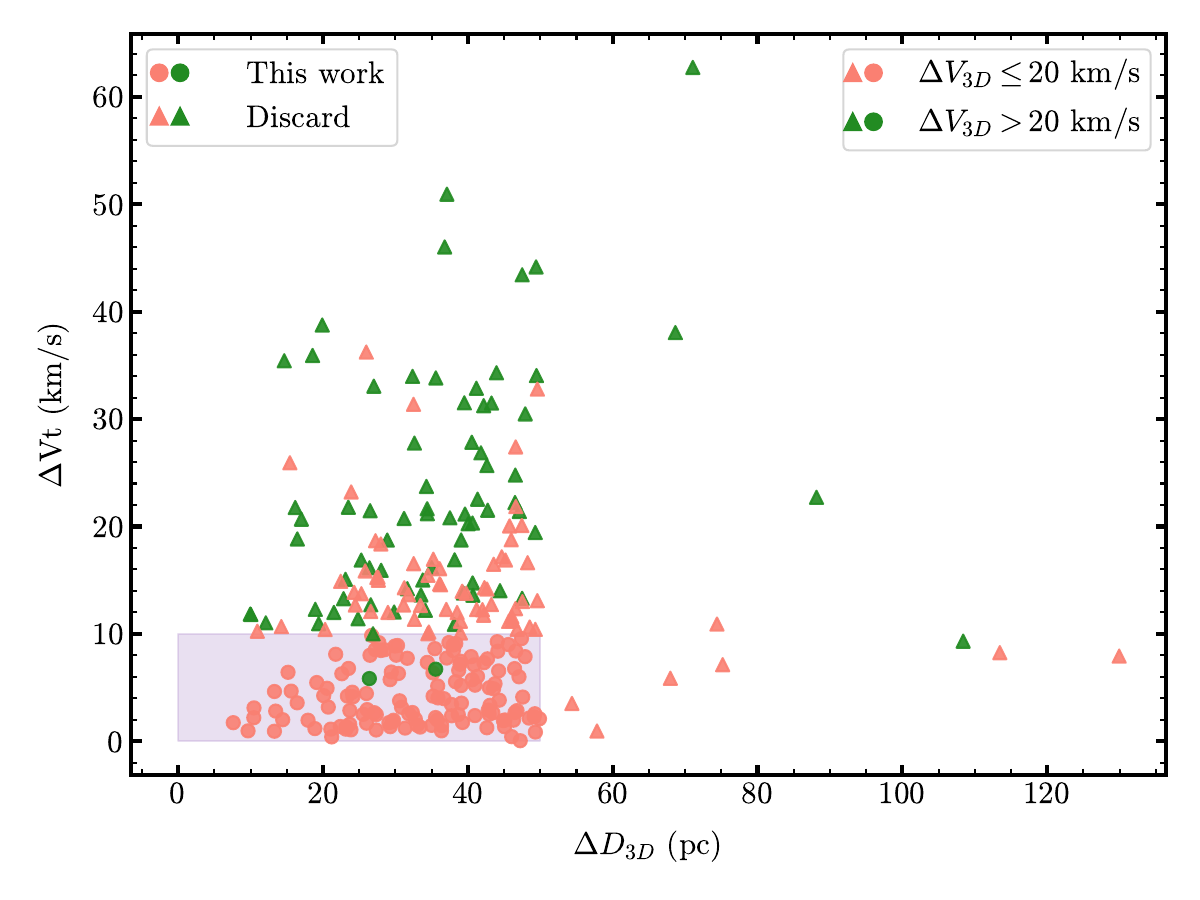}
    \caption{Spatial separation as a function of tangential velocity for all matched cluster pairs from the literature. Circles denote BC candidates confirmed in this work, while triangles represent discarded pairs. Red symbols represent pairs with $\Delta V_{3D} \leq 20$ \kms, and green symbols indicate those with $\Delta V_{3D} > 20$ \kms. The shaded region marks the selection range adopted for our BC sample.}
    \label{fig:compare_lit}
\end{figure}

\begin{table*}[htbp]
\centering
\caption{Open cluster group candidates.}
\setlength{\tabcolsep}{5.0pt}
\renewcommand{\arraystretch}{1.5}
\begin{tabular}{rrrrrrrrrrrrr}
\hline \hline 
Groups & Pairs & $\text{ID}_\text{HR24}$ & Cluster & \RAdeg & \DEdeg & \plx & \pmra & \pmdec & RV & logt & DM & $E(B-V)$ \\ 
{} & {} & {} &{} & deg & deg & mas & \masyr & \masyr &\kms & {}&mag&mag\\
(1) & (2) & (3) & (4) & (5) & (6) & (7) & (8) & (9) & (10) & (11) & (12) & (13) \\
\hline
1 & 5 & 6925 & UBC\_17a & 84.78 & $-$1.93 & 2.77 & 1.80 & $-$1.30 & 11.74 & 6.50 & 8.58 & 0.13 \\ 
1 & 320 & 5102 & OCSN\_61 & 84.08 & $-$0.40 & 2.59 & $-$1.04 & $-$0.63 & 26.82 & 7.43 & 7.53 & 0.01 \\ 
1 & 5 & 11 & ASCC\_19 & 81.98 & $-$1.86 & 2.81 & 1.17 & $-$1.18 & 11.65 & 7.11 & 7.91 & 0.11 \\ 
2 & 2 & 7 & ASCC\_12 & 72.42 & 41.71 & 0.95 & $-$0.68 & $-$2.92 & $-$26.14 & 8.02 & 11.09 & 0.33 \\ 
2 & 2 & 4203 & HXHWL\_18 & 74.32 & 39.80 & 0.96 & $-$0.53 & $-$3.06 & $-$9.60 & 7.55 & 10.68 & 0.31 \\ 
2 & 190 & 2851 & HSC\_1308 & 76.51 & 38.47 & 0.95 & $-$0.36 & $-$3.13 & $-$16.74 & 7.51 & 11.40 & 0.55 \\ 
2 & 40 & 326 & COIN-Gaia\_39 & 69.58 & 42.97 & 0.97 & 0.10 & $-$2.54 & 7.96 & 8.31 & 10.75 & 0.34 \\ 
2 & 40 & 300 & COIN-Gaia\_10 & 68.46 & 40.50 & 0.95 & 1.97 & $-$3.45 & 8.09 & 7.88 & 11.50 & 0.53 \\ 
... &... & ... & ... &... &... & ... & ...  & ... &... & ... & ...  & ... \\ 
\hline
\end{tabular}
\tablefoot{Column (1): Group IDs. Column (2): Pair IDs. Column (3): Cluster IDs from \citealt{Hunt2024}. Column (3): Cluster names. Columns (5)-(10) are adopted from HR24, with all values referenced to epoch J2016.0. Columns (11)-(13) are derived in this work, see also \tabref{tab: bc_candidate}. The full Table is available at the CDS. Here we only show the first two OC groups.}
\label{tab: group_candidate}
\end{table*}

\subsection{Compare with literature}
Cross-matching our identified cluster pairs with previously reported BCs provides a powerful way to validate the effectiveness of our identification methodology. The most famous BC is the $h$ and $\chi$ Persei pair (NGC 869 and NGC 884; \citealt{Messow1913, Zhong2019A&A}), which serves as a valuable benchmark for testing identification results. However, this pair is not included in our golden sample because their mean RV are based on only two and six member stars, respectively, making the measurements less reliable. Nevertheless, they are included in our full sample, suggesting that using $V_t$ is effective for identifying BCs when reliable RVs are unavailable.

Building on this case, we expanded our validation by performing a comprehensive cross-match with previously reported BC samples. We first compiled the sample from the literature, which includes nearly 800 pairs listed in \tabref{tab: literature}.
These catalogs in \tabref{tab: literature} encompass diverse datasets and reflect the evolving approaches to identifying and analysing these systems. We then cross-matched these literature samples with our BCs. We found that 230 pairs show partial overlap with literature-reported BCs, among which 132 pairs are in exact agreement. 

\figref{fig: lit_repeat} summarizes the overlap between cluster pairs identified in this work and those reported in the literature, categorized by BC type. Among the 243 PBCs, 70 match previously known systems, while 173 (71.2\%) are newly identified. For the 146 TBCs, 60 are duplicates and 86 are new. In the HEP category, only 2 of the 11 systems were previously reported as BCs. These results confirm that our method is capable of recovering known binary systems with high reliability while also discovering a large number of new physically associated pairs, particularly in the primordial BCs. This significantly expands the available BC sample in the Milky Way.

To comprehensively evaluate the coverage of literature samples, we compared the 800 previously reported BC pairs with our set of 2,974 nearest-neighbor OC pairs.  We found that 1,123 of our pairs show partial overlap with literature-reported systems, among which 275 pairs are exact matches. \figref{fig:compare_lit} shows the distributions of $\Delta D_{3D}$, $\Delta V_t$, and $\Delta V_{3D}$ for these 275 pairs. It can be seen that the matched candidates not included in our BC sample have larger velocity differences, highlighting differences in the velocity selection criteria used and the impact of our stricter kinematic selection criteria.

Among the overlapping systems, 252 pairs originate from the catalog of P25, representing the largest source of overlap. Accordingly, this dataset serves as the primary basis for comparison. P25 constructed a catalog of 617 cluster pairs using the same HR23 and HR24 data but adopted a different selection approach, which contains 53 genetic BCs, 88 formed through tidal capture or resonant trapping, and 476 classified as optical pairs, likely arising from gravitational (hyperbolic) encounters. 

The discrepancy between our results and those of P25 primarily arises from differences in both the OC sample and selection criteria. Although both studies utilize data from HR23 and HR24, our search is based on the high-quality subset of star clusters defined in HR24. In contrast, P25 includes 7,046 entries from the HR24 catalog. 
As a result, the nearest neighbor of a given OC may differ from that in previous works, leading to reassignments and only partial overlap with earlier catalogs. For example, some pairs previously identified as BCs may now be re-matched with newly added, closer neighbors in our sample.
Notably, many of these low-quality star clusters are not particularly scientifically useful for studies of star clusters, as they cannot be validated as real with currently available data \citep{Hunt2024}. 

Despite these differences, the overlap with previously reported systems provides strong support for the reliability and effectiveness of our identification method. 
This result demonstrates that our method is effective in identifying previously reported BCs.  Our work not only recovers a large fraction of previously reported systems but also contributes 268 newly identified physical BCs to the Galactic sample.

Based on our results, approximately 16.80\% of OCs appear to be experiencing some type of interaction with a companion cluster, and 9.94\% likely formed as PBCs, which formed together in the same GMCs.These fractions are in good agreement with previous theoretical and observational estimates of 8-20\% for the BC occurrence rate in the Milky Way \citep[e.g.,][]{Subramaniam1995, Priyatikanto2019, delaFuenteMarcos2009a, Casado2021b}.

\begin{figure}[htbp]
    \centering
    \includegraphics[width=1\linewidth]{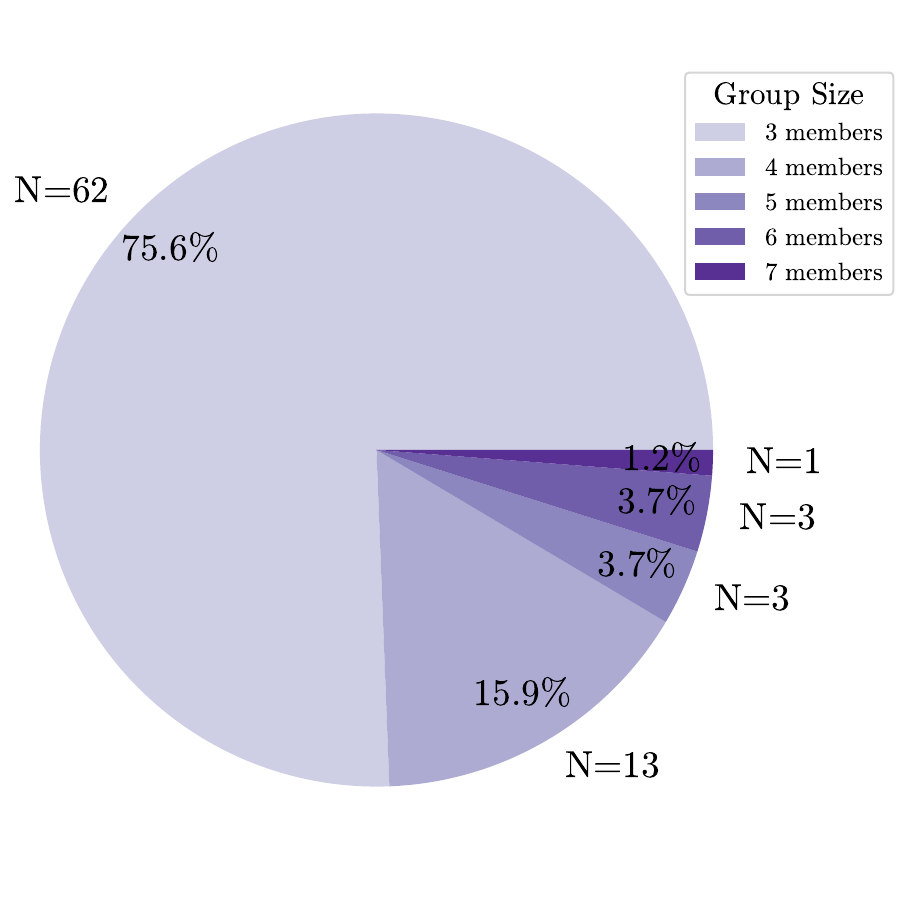}
    \caption{The pie chart shows the number of star clusters in each group. Colors are coded by the number of group members.}
    \label{fig: group_count}
\end{figure}

\subsection{OC groups}
Based on the BC sample, we identified OC groups using the principle of transitivity, following the method introduced by \citet{Piecka21}. Specifically, if Cluster A and Cluster B form a BC, and Cluster B and Cluster C also form a BC, then Clusters A, B, and C are considered members of the same group.
Applying this transitive criterion to our BC sample, we identified a total of 278 clusters involved in multi-cluster systems, resulting in 82 distinct OC groups. Among these, 60 are triple-cluster systems, 12 are quadruple systems, three are quintuple systems, three are sextuple systems, and one is a septuple system. The results are shown in \figref{fig: group_count}, and the properties of all OC groups are summarized in \tabref{tab: group_candidate}.

To assess the novelty and completeness of our group catalog, we cross-matched our identified OC groups with approximately 330  previously reported groups, which are compiled in \tabref{tab: literature}. The results of this comparison are summarized as follows: \\
$-$ Complete matches were found for 29 groups, including 28 from P25 and 1 from \citet{Liu2019}. These overlapping systems mostly consist of small groups: 22 triple-cluster systems, 4 quadruple systems, and 1 sextuple system.\\
$-$ Subset matches were identified in 26 additional groups. In these cases, one group is entirely contained within the other, meaning that either a literature-reported group is fully embedded in a larger group from our catalog, or our identified group is entirely included within a larger group reported in the literature.\\
After excluding these 55 overlapping cases, our study reports 27 newly identified OC groups.
These systems provide further support for the hierarchical star formation scenario, extending it from small-scale OCs or BCs to larger-scale structures \citep{Elmegreen1996}.

\section{Summary} \label{sec: summary}
In this study, we present a comprehensive search and classification of BCs in the Milky Way, based on a high-quality sample of 4,084 star clusters from the HR24 catalog.
To define robust identification criteria, we analyzed the nearest-neighbor distributions of spatial and velocity separations among OCs and constructed mock datasets to assess the statistical significance of clustering. These tests support selection thresholds of $\Delta D_{3D} \leq$ 50 pc and $\Delta V_{3D} \leq$ 20 \kms. To further include pairs lacking reliable radial velocities and also constrain the velocity vector, we adopted a $\Delta V_t$ threshold of 10 \kms, following previous studies. 
Based on these criteria, we first identified 400 candidate BCs with $\Delta D_{3D} \leq$ 50 pc and $\Delta V_t \leq$ 10 \kms, which we define as the full sample. Among them, 146 pairs also meet the 3D velocity criterion and constitute the golden sample.

We classify 400 BC candidates into three types based on age and kinematic consistency. There are 243 PBCs, accounting for 60.8\% of the sample. These PBCs have similar ages and motions, suggesting the member OCs formed together in the same GMC. Another 146 pairs are TBCs, with similar motions but different ages, likely formed through a sequential mode. The remaining 11 pairs are HEPs. They are close in space but have large velocity differences, indicating they are not born physically bound.

We further calculate the TF for each pair and find that 82.5\% of our BCs satisfy TF $<$ 200, reinforcing the physical association of the majority of the sample. Additionally, we identify 278 clusters belonging to 82 OC groups, suggesting a hierarchical formation structure.
In addition, 278 star clusters are identified as members of 82 multi-cluster systems, including 27 newly reported OC groups. These groups exemplify the hierarchical star formation scenario from small-scale BCs to larger-scale structures. 

Cross-matching with about 800 BCs reported in the literature confirms that our methodology recovers 132 previously known pairs and contributes 268 newly identified systems. We estimate that 16.80\% of OCs are currently part of BC or group systems, and 9.94\% likely formed as PBCs. These values are consistent with prior theoretical predictions and observational estimates.
Our work provides a homogeneous and statistically robust catalog of BCs, offering a foundation dataset for future studies of hierarchical star formation, stellar feedback, and dynamical evolution within the Galactic disk.

\section{Data availability}
Tables 2 and 3 are only available in electronic form at the CDS via anonymous ftp to cdsarc.u-strasbg.fr (130.79.128.5) or via http://cdsweb.u-strasbg.fr/cgi-bin/qcat?J/A+A/.

\begin{acknowledgements}
We are grateful to the anonymous referee for valuable comments that significantly improved the quality of this paper. We also thank Shiyin Shen and Ruqiu Lin for their insightful discussions and constructive suggestions.

Yu Zhang acknowledges the support from the science research grants from the Chinese Academy of Sciences (CAS) "Light of West China" Program (No. 2022-XBQNXZ-013), Central Guidance for Local Science and Technology Development Fund (No. ZYYD2025QY27), and Natural Science Foundation of Xinjiang Uygur Autonomous Region (No. 2022D01E86).

This work is supported by the National Natural Science Foundation of China (NSFC) through the grants 12090040 and 12090042.
Jing Zhong would like to acknowledge the science research grants from the China Manned Space Project with NO. CMS-CSST-2025-A19, the Youth Innovation Promotion Association CAS, the Science and Technology Commission of Shanghai Municipality (Grant No.22dz1202400), and the Program of Shanghai Academic/Technology Research Leader.

Songmei Qin and Yueyue Jiang acknowledge financial support from the China Scholarship Council (Grant Nos. 202304910547 and 202404910435, respectively).

\\
This work has made use of data from the European Space Agency (ESA) mission \gaia \url{https://www.cosmos.esa.int/ gaia}, processed by the Gaia Data Proces-Processingalysis Consortium (DPAC,\url{https://www.cosmos. esa.int/web/gaia/dpac/consortium}). Funding for the DPAC has been provided by national institutions, in particular, the institutions participating in the Gaia Multilateral Agreement.
\end{acknowledgements}

\bibliographystyle{aa}
\bibliography{reference}
\end{CJK*}
\end{document}